\documentclass[12pt,runningheads,a4paper]{llncs}

\usepackage{times}
\usepackage{epsfig}
\usepackage{graphicx}
\usepackage{amsmath}
\usepackage{amssymb}
\usepackage{enumitem}
\usepackage[hyphens]{url}
\newenvironment{packed_item}{
	\begin{itemize}
		\setlength{\itemsep}{0pt}
		\setlength{\parskip}{0pt}
		\setlength{\parsep}{0pt}
		\setlength{\leftmargin}{0pt}
	}{\end{itemize}}

\newenvironment{packed_enum}{
	\begin{enumerate}
		\setlength{\itemsep}{0pt}
		\setlength{\parskip}{0pt}
		\setlength{\parsep}{0pt}
		\setlength{\leftmargin}{0pt}
	}{\end{enumerate}}

\newenvironment{packed_enumc}{
	\begin{enumerate}[label=Step \arabic*,leftmargin=*,labelindent=10pt]
		\setlength{\itemsep}{0pt}
		\setlength{\parskip}{0pt}
		\setlength{\parsep}{0pt}
		\setlength{\leftmargin}{0pt}
	}{\end{enumerate}}

\newcommand{\Figref}[1]{Fig.~\ref{#1}}
\begin{document}

\mainmatter 

\title{Access Control in Internet of Things: A Survey}

\titlerunning{Access Control in Internet of Things: A Survey}

\author{Yunpeng Zhang
\and
Xuqing Wu
}

\authorrunning{Z. Yunpeng and X. Wu}

\institute{Department of Information and Logistics Technology,\\
	College of Technology, University of Houston,\\
	Houston, TX 77204\\
}

\maketitle

\begin{abstract}
Cheating is a real problem in the Internet of Things. The fundamental question that needs to be answered is how we can trust the validity of the data being generated in the first place. The problem, however, isn’t inherent in whether or not to embrace the idea of an open platform and open-source software, but to establish a methodology to verify the trustworthiness and control any access. This paper focuses on building an access control model and system based on trust computing. This is a new field of access control techniques which includes Access Control, Trust Computing, Internet of Things, network attacks, and cheating technologies. Nevertheless, the target access control systems can be very complex to manage. This paper presents an overview of the existing work on trust computing, access control models and systems in IoT. It not only summarizes the latest research progress, but also provides an understanding of the limitations and open issues of the existing work. It is expected to provide useful guidelines for future research.
   
   \vspace{2.0ex}
   \noindent Access Control, Trust Management, Internet of Things
\end{abstract}


Today, our world is characterized by increasing connectivity. ‘Things’ in this world are increasingly being connected. Smart phones have started an era of global proliferation and rapid consumerization of smart devices. It is predicted that the next disruptive transformation will be the concept of `Internet of Things'~\cite{atzori2010}. From networked computers to smart devices, and to connected people, we are now moving towards connected `things'. Items of daily use are being turned into smart devices as various sensors are embedded in consumer and enterprise equipment, industrial and household appliances and personal devices. Pervasive connectivity mechanisms build bridges between our clothing and vehicles. Interaction among these things/devices can happen with little or no human intervention, thereby conjuring an enormous network, namely the Internet of Things (IoT). One of the primary goals behind IoT is to sense and send data over remote locations to enable detection of significant events, and take relevant actions sooner rather than later~\cite{nukala2012}.

This technological trend is being pursued actively in all areas including the medical and health care fields. IoT provides opportunities to dramatically improve many medical applications, such as glucose level sensing, remote health monitoring (e.g. electrocardiogram, blood pressure, body temperature, and oxygen saturation monitoring, etc), rehabilitation systems, medication management, and ambient assisted living systems. The connectivity offered by IoT extends from human-to-machine to machine-to-machine communications. The interconnected devices collect all kinds of data about patients. Intelligent and ubiquitous services can then be built upon the useful information extracted from the data. During the data aggregation, fusion, and analysis processes, user privacy and information security become major concerns for IoT services and applications. Security breaches will seriously compromise user acceptance and consumption on IoT applications in the medical and health care areas.

The large scale of integration of heterogeneous devices in IoT poses a great challenge for the provision of standard security services. Many IoT devices are vulnerable to attacks since no high-level intelligence can be enabled on these passive devices~\cite{gubbi2013internet}, and security vulnerabilities in products uncovered by researchers have spread from cars~\cite{humayed2015} to garage doors~\cite{ANDY15} and to skateboards~\cite{kim15}. Technological utopianism surrounding IoT was very real until the emergence of the Volkswagen emissions scandal~\cite{Burki2015}. The German conglomerate admitted installing software in its diesel cars that recognizes and identifies patterns when vehicles are being tested for nitrogen oxide emissions and cuts them so that they fall within the limits prescribed by US regulators ($0·04$ g/km). Once the test is over, the car returns to its normal state: emitting nitrogen oxides (nitric oxide and nitrogen dioxide) at up to 35 times the US legal limit. The focus of IoT is not the thing itself, but the data generated by the devices and the value therein. What Volkswagen has brought to light goes far beyond protecting data and privacy, preventing intrusion, and keeping the integrity of the data. It casts doubts on the credibility of the IoT industry and its ability to secure data, reach agreement on standards, or indeed guarantee that consumer privacy rights are upheld.	
	
All in all, IoT holds tremendous potential to improve our health, make our envi- ronment safer, boost productivity and efficiency, and conserve both water and energy. IoT needs to improve its trustworthiness, however, before it can be used to solve challenging economic and environmental problems tied to our social lives. The fundamental question that needs to be answered is how we can trust the validity of the data being generated in the first place. If a node of IoT cheats, how does a system identify the cheating node and prevent a malicious attack from misbehaving nodes? This paper focuses on an access control mechanism that will only grant network access permission to trustworthy nodes. Embedding trust management into access control will improve the system’s ability to discover untrustworthy participating nodes and prevent discriminatory attacks. There has been substantial research in this domain, most of which has been related to attacks like self-promotion and ballot stuffing where a node falsely promotes its importance and boosts the reputation of a malicious node (by providing good recommendations) to engage in a collusion-style attack. The traditional trust computation model is inefficient in differentiating a participant object in IoT, which is designed to win trust by cheating. In particular, the trust computation model will fail when a malicious node intelligently adjusts its behavior to hide its defect and obtain a higher trust value for its own gain.
	
	\section{Access Control Model and System}
	
	IoT comprises the following three Access Control types：
	\begin{packed_item}	
		\item Role-based access control (RBAC)
		\item Credential-based access control  (CBAC) --- in order to access some resources and data, users require certain certificate information that falls into the following two types:
		\begin{packed_enum}
			\item Attribute-Based access control  (ABAC) : If a user has some special attributes, it is possible to access a particular resource or piece of data.
			\item Capability-Based access control  (Cap-BAC): A capability is a communicable, unforgeable rights markup, which corresponds to a value that uniquely specifies certain access rights to objects owned by subjects.
		\end{packed_enum}
		\item Trust-based access control (TBAC)
	\end{packed_item}
	
	In addition, there are also combinations of the aforementioned three methods. In order to improve the security of the system, some of the access control methods include encryption and key management mechanisms.
	
	\subsection{RBAC}
	In the case of an emergency (i.e. a doctor is urgently needed after a traffic accident), the user's location should be accessible (under normal circumstances, information about the user's location that impinges on personal privacy should be kept confidential). Reference\cite{hu2011identity} proposes an identity-based system to manage personal location information in emergencies. It includes registration, user authentication, policy, and a client terminal system. The system authenticates the user's identity through a user authentication subsystem; it also obtains an emergency rating by means of a policy subsystem. This ensures only certain authorized users can access a user's location.

 In reference \cite{liu2012authentication}, Liu, et al., propose authentication and access control methods for IoT. The authors analyze the existing authentication and access control and design a realizable access control protocol for IoT. The protocol focuses on the process of establishing a simple and efficient ECC-based security key. In the access control policy, the authors use an RBAC authorization method based on the requesters' specific roles and applications related to the IoT network. However, due to the highly dynamic environment and the huge number of users of IoT, RBAC cannot assign permissions in advance, and the method places a high toll on perception nodes in the entire communication process. In addition, the reliability evaluation in the actual situation is not applicable. Thus, reference \cite{ndibanje2014security} modifies the protocol at the expense of safety and operational aspects based on reference \cite{liu2012authentication}, and analyzes the performance of an improved protocol. The new protocol includes the registration phase (offline or online), login and authentication phase. In addition, the new protocol also incorporates password recovery and modification capabilities to help users manage passwords. In the registration phase, each user needs to be registered with the main registry, resulting in negotiation and calculation of the key parameters between the user and the gateway node in the login and authentication phases, thereby completing the mutual authentication between nodes in two phases of the login and authentication. The analysis shows that the new method can meet the confidentiality, reliability, integrity and other key security requirements. However, due to the huge number of nodes and dynamic environment of IoT and limited computing and storage capacity of an IoT node, the applicability of these types of methods will be greatly limited.	
	
	\subsection{CBAC}
	
\noindent \textbf{ABAC}	Traditional access control models, such as ACLs (Access Control List), RBAC, and ABAC, due to their inherent limitations, cannot be directly applied to IoT systems. Due to the characteristics of ABAC, it has a great advantage and has been widely used in the traditional Internet. However, it cannot be directly applied to IoT because of limitations of flexibility. Existing research papers mainly combine ABAC with other methods to obtain suitable access control methods for IoT. Due to the highly dynamic environment and the huge number of users of IoT, RBAC cannot assign permissions in advance with traditional access control methods. Moreover, the ABAC authorization process is complex: it does not apply to the highly dynamic, real-time environment of IoT and the number of rules rapidly increases with users, attributes and growth. Nevertheless, RBAC and ABAC still have some advantages that can be exploited. RBAC can effectively solve the distribution problem of competencies with time and location changes, while ABAC can solve the dynamic propagation problems of users. Therefore, reference \cite{kaiwen2014attribute} proposes a hybrid access control model based on role and attribute (ARBHAC), which can resolve the large-scale dynamic problem of users in IoT. This model pre-assigns roles for nodes/users based on the property expression of nodes/users, and then assigns roles distribution and corresponding permissions to the appropriate nodes/users. The model presents a property rule policy language and a solution to the conflict with the redundancy policy. The authors use the Weichat App as an example to illustrate the feasibility of this model. This model simplifies the complexity of traditional ABAC in rights allocation and policy management. However, it is not enough to deal with the policy conflict and redundancy processing as the model still needs the administrator to manage roles and permissions licenses. In short, this method requires further research to improve its usability. Attribute-based encryption policy ciphertext (CP-ABE) is a powerful and asymmetric encryption mechanism, which allows fine-grained access control. The lack of a valid key in the current regime of the CP-ABE revocation management mechanism is the main shortcoming. Researchers found solutions for this shortcoming. In some of the existing solutions, there is a problem in the transmission and consumption of complexity. The other solutions require a strong trust agent to decrypt the data. Reference \cite{touati2015activity} proposes access control based on activities (Activity-based Access Control), which is a generalized version of context-aware, an account of the user context changes, with fine-grained features. Its aim is to make decisions based on perceived user stories and system status. By using a finite state ma- chine and a CP-ABE (Ciphertext-Policy Attribute-Based Encryption) which is an asymmetric encryption mechanism, a real-time access control policy can be adapted according to situational changes in the user and system. A finite state machine also can be used to simulate a user state, where each object has a unique association with a finite state machine, which is constructed from the system design phase. The Attribute Authority is used for storage. The finite state machine contains all possible states of the system life cycle of the object and specifies the object state changes based on predefined time and manner. Each state corresponds to a set of access privileges. The program also takes into account the real-time properties of the Revocation mechanism. Reference \cite{touati2015batch} proposes a revocation mechanism called attribute-based CP-ABE, which is based on a batch method and can reduce processing complexity and overall consumption and does not need to add additional nodes in the system. In this paper, the timeline is divided into equally spaced intervals: access policy changes occur only on two consecutive time slots and it is only necessary to send one key for updating. So the solution in this paper can achieve the purpose of revocation. To maximize system performance and minimize the average waiting time, the authors also analyze the chosen method of the optimal time slot construction. Reference \cite{lang2015proximity} focuses on Proximity-based Access Control (PBAC), which is a relatively new method of access control. It is a particularly advanced access control approach that can implement flexible, proximity-based, dynamic, contextual access. PBAC, ABAC and MDS (Model-Driven Security) are all for expression and implementation of security and privacy requirements. PBAC uses the policy based on proximity with requesters and resources. The relationship is not limited to the physical proximity of adjacent elements. It is further composed of a number of adjacent elements: geographic location, organization, operation, business process, security risk, social factors, and other information. PBAC has to include an adjacent calculation function which indicates the relationship between these close properties. An extension of the ABAC is necessary to support fine-grained, flexible, context-sensitive access control policies based on proximity. In addition, the use of an MDS mechanism can be implemented to generate MDS and MDS certification automation strategies. So ABAC will be more manageable and the PBAC policy will be safer. The paper also describes the policy management approach by using MDS and extended ABAC with a detailed example involving intelligent transport systems.

\subsection{Cap-BAC} 
Reference \cite{accessLink} proposes an access control model (Cap-BAC) based on existing abilities. This model may be based on the principle of least privilege to manage access to services and information control processes. In Cap-BAC, the user needs to show the service provider the authorization certificate prior to performing corresponding resource request operations. Authorization certificates are issued by the owner of a resource/service to desired users, in order to ensure that the users can request resources or services. It emphasizes the availability of a security mechanism and correlation between access authorization in addition to the need for methods that are understandable and accessible (especially for users with limited knowledge of information and communication). The paper acknowledges the default principle of least privilege, supports revocation of privilege and control of validity for a certificate of competency, providing adequate security and flexibility for the practical application of the program. The main disadvantage of this solution is that it requires the ability to publish all the main certificates (although a privilege delegation mechanism can solve part of this problem) and to have the selection capability available when a certificate body submits a request. To solve this problem, authorized policy ability can be used to set a specific service and then to generate the appropriate access rights for recognized and authorized users. In addition, because other access control mechanisms must be developed under cross-domain or cross-enterprise environmental performance, they require the ability to standardize the structure of the certificate, whereas Cap-BAC's strongpoints are support services and an access control protocol. Reference \cite{sicari2014nfp} defines a universal UML concept model that can be used for all IoT applications and architectures. It specifies the entities involved in IoT infrastructure and their relationships, as well as pointing out their roles and functions. The model takes into account the stage where users register with the IoT platform in the handling of personal information and certificate exchanges in future interactions. It provides a new way of thinking about the management of registered users, things, and relevant certificates; however, there is still the need to establish a standardized and universally-accepted solution. Reference \cite{gonccalves2013security} provides a basic security architecture in which secure support and verifiable interactions can easily be deployed on a mobile platform. The Transport Layer Secure Sockets Layer (SSL) protocol is proposed for M-Health Security Protocol (MHSP), which focuses on the application layer protocol for authentication between the network and the application. It can provide a secure channel, using the public key certificate (certificate properties include: device identification and the MAC address of the network drivers) for authentication review. Through the use of personal electronic health records, this architecture can establish and manage prescription services in a mobile environment. Relying on RFID technology and a simple and intuitive interface, the architecture can be used for mobile e-health applications (M-Health). It is also joined together with a corresponding IoT context, for drug supervision management, medication control systems under special circumstances, analysis of the proposed security architecture and protocols, as well as medications to control ambient assisted living services (AALSS) in an IoT environment.

\subsection{TBAC}
IoT needs a flexible, lightweight and adaptive access control mechanism to deal with its universal nature and to ensure credible and reliable communication be- tween devices. Reference \cite{bernabe2015taciot} proposes the flexible trust perceptual characteristics of the access control system for IoT (TACIoT). It is a lightweight authorization mechanism designed specifically for a networking model based on trust to provide a reliable security mechanism from beginning to end in IoT. TACIoT extends the traditional access control system and includes the value of trust based on reputation, quality of service, security considerations and social equipment. TACIoT has been successfully implemented in a practical experimental stage, which was evaluated using constrained and unconstrained networking equipment. Reference \cite{mahalle2013fuzzy} indicates that the traditional access control model is not suitable for a distributed dynamic IoT environment because the identity of devices cannot be learned in advance. A trust relationship between two devices will directly affect the interactions between them. When two devices trust each other, they will be more willing to share services and resources. This paper proposes a method of access control (FTBAC) with a fuzzy trust value. The trust value between devices is calculated based on the framework FTBAC --- Experience, Knowledge, and Recommendation, among other factors. The trust value can be mapped to the appropriate authority. The set of certificates and access requests are the access credentials. The FTBAC framework consists of three layers: 1) Device Layer: includes all IoT devices and their communications; 2) Requesting layer: mainly responsible for collecting EKR (Enterprise Knowledge Repository) information and calculating the fuzzy trust value; 3) Access Control layer: includes the decision-making and mapping process of the fuzzy trust value and access
based on the principle of least privilege. The simulation results show that the frame- work can ensure flexibility, scalability, and more energy. In fact, a solution based on encryption-protection can gain access to the controls by increasing the level of trust, but it requires additional time and energy consumption. Moreover, the fuzzy method is more readily combined with utility-based decision making.

\subsection{IBAC}
\begin{figure}[htbp]
	\centering
	\begin{tabular}{c}
		\includegraphics[width=0.8\textwidth]{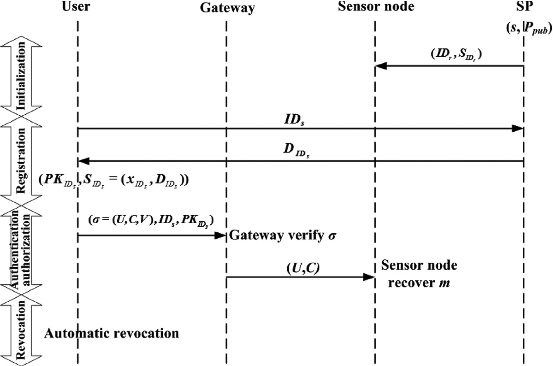}
	\end{tabular}
	\caption{An access control scheme}
	\label{fig1}
\end{figure}

The IBAC is a simple and practical access control model that associates access privilege with specific users. The pro-posed scheme consists of four phases: the initialization phase, the registration phase, the authentication and authorization phase, and the revocation phase. In this scheme, the SP plays the role of PKG in the IBC environment and the KGC in the CLC environment. The proposed access control scheme is summarized in ~\Figref{fig1}~\cite{li2016practical}.
	
\subsection{DCapBAC}	
DCapBAC~\cite{hernandez2016dcapbac} allows a distributed approach in which constrained devices are enabled with access control logic by considering suitable technologies and mechanisms for IoT environments. The capability-based access control system is being deployed and tested in SOCIoTal scenarios as key part of the Authorization Systemof security framework. DCapBAC also focuses on the authorization enforcement stage, whereby the involved devices can be mutually authenticated and the capability tokens are exchanged securely and then validated for a specific access request.	
	
\subsection{RFID access control}
Reference \cite{jensen2013access}  proposes a method to incorporate guest objects/things into IoT. The authors constructed an RFID-based access control system based on packet filtering and network layer technology. The method uses low-cost, passive RFID tags attached to objects in a virtualized representation, giving identity to the objects in IoT. Access control rights management in the Internet layer is analogous to a packet filtering firewall, so the authors built an access control system based on packet filtering between the Internet and object layers. The system constructs a prototype RFID tag which is mapped to IPv6 addresses. RFID tags are conversed to a network layer identity by assigning IP addresses to RFID. The scheme uses network technology to replace the access control list of the RFID access control system to complete the communication system in the distribution of components. By using the packet filtering IPv6 addressing scheme and organized unauthorized access requests on an access control node, the method does not directly interact with the back-end database systems in the process. The huge address space of IPv6 can meet the huge amount of reader and tag entity needs. Through this example, the authors explain how the method is applied to the access control system which is based on open network protocol and packet filtering. The method includes a new RFID reader architecture to support future network-based access control system components based on networking layer technology. A reading device acts as a standalone access control system. It includes independent storage access control rules. Only when the rule needs to be updated, do reading devices need to communicate with the server. Rules can be updated by the multicast method. In the same security zone, multiple reading devices can distribute safety rules at the same time, thereby improving the efficiency of rule updates.
	
\subsection{Key management in Access Control System}	
Reference \cite{cherkaoui2014new} focuses on authentication and access control within the IoT framework, proposing authorization schemes for constrained devices including the physical unclonable function (PUF) and embedded subscriber identity module (eSIM). PUF provides an inexpensive, safe, tamper-resistant key to verify constrained M2M equipment, and eSIM provides scalable mobile connectivity management, interoperability, and compliance with security protocols. Reference \cite{veltri2013novel} describes the use of a shared key to multiple communication terminals (denoted as the group key) as the method to protect the security of the multiplex. The key consists of a centralized approach to management and distribution of the batch. The program can reduce computing consumption and network transmission resulting from changes in group membership as users come and go from the network. Reference \cite{majun2013iot} focuses on access control for a data acquisition layer. In accordance with the level of privacy and security, the layer requires a significant amount of nodes for authorized user perception encompassing different data types. Reference \cite{majun2013iot} proposes a hierarchical access control scheme. Taking into account the limited computing and storage capacity of the nodes, the scheme provides each user and node with a single key and then uses the confirmed key generation algorithm to calculate additional keys as needed. Since key exchange is limited, this scheme can improve safety and reduce storage consumption by the nodes. 

Reference \cite{paek2015fast} develops a fast adaptive network access control method for low-energy lossy networks (LLNs). LLNs contain thousands of embedded network devices, which are connected to a large network architecture. They can be used for various applications and produce a new concept of IoT. In such LLNs, access control and key management are very important issues because the authentication, authorization, and key management process includes a plurality of large data packets in low bandwidth and high latency connections during the exchange process. A bottleneck effect at the router border may lead to a significant waste of bandwidth and slower completion of identity verification. This paper proposes a``low-energy lossy and fast adaptive network access control'' method called FINALLY, which allows the device to authenticate once after a failed attempt and a custom-selected waiting time. It can eliminate unnecessary waste of bandwidth and latency. With the reduced waiting time, it can create a linear arrangement for every terminal space in time, thereby avoiding congestion and unnecessary delays for the entire network. It can solve these problems and significantly improve the efficiency and speed of the network authentication process. Simulated evaluation results show that it can guarantee fast authentication devices and avoid repeated validation failure, thereby reducing unnecessary consumption of transmission, and improving the overall completion time of the verification process. Reference~\cite{di2016programmable} propose a novel key techniques for multi-hop wireless networks for IoT. This key techniques include three parts, network tomography, network division and traffic scheduling, which support the network resources optimally and dynamically managed among these networks to meet all application requirements.	

\subsection{Real-time transport control data collection}
Real-time information collection with IoT in the external environment is more convenient and more effective. But when we apply the traditional access control scheme in network communications, it will cause serious traffic congestion. Reference \cite{kawamoto2015novel} proposes a new IoT in the outside world of information access control mechanisms to build a real-time database, which can effectively avoid communication congestion. In addition, it provides mathematical methods to improve the efficiency of the optimization method. The mechanism is set dynamically according to the change rate of the transmission time of each of the terminals on the external environment. To set a message transmission time, the terminal needs to periodically check the current information on the outside of the terminal and be set to a rate of change between the information servers. When the change rate is higher than a predetermined threshold, the execution information transmission process begins. In order to optimize the process, ensure a real-time database and avoid blocking the transmission case, the authors also provide a mathematically optimal method of setting the threshold value.

	\section{Trust computing}

	
	Trust computation model and trust management systems have been implemented successfully in commercial applications~\cite{yan2014survey,josang2007survey}. There is also a rapidly growing literature around topics of trust and reputation management for IoT. The motivation of initiating a trust computation model is to prevent discriminatory or malicious attacks from misbehaving nodes. In addition, misbehaving nodes may comprise the integrity and quality of services provided by IoT devices in an integrated environment for the medical and health care. Therefore, it is necessary to establish trustworthiness among devices. The strategies of building a reputation-based trust mechanism for IoT health care devices should deal effectively with a certain type of malicious behavior that intends to mislead other nodes. According to \cite{chen2015trust}, trust-related attacks include: (1) self-promoting attacks, which promote own credibility through illegal means; (2) bad-mouthing attacks: which reduce the trust value of good nodes; (3) ballot-stuffing attacks, which boost the reputation of malicious nodes; (4) opportunistic service attacks, which raise their own reputation through providing quality service in a random manner; (5) on-off attacks, which provide poor services intermittently. 
	
	Recent work in conducting the survey for most relevant available solutions to trust management, models, and frameworks include \cite{josang2007survey,yan2014survey,sicari2015security}. The main motivation of this paper is to identify trust management frameworks that are suitable for IoT applications in the area of medical and health care. Especially, we are interested in trust value computation and trust value propagation in different the types of the frameworks.

	Sun et al.~\cite{sun2006information} suggested that trust is a relationship established between two entities for a specific action. The also introduced a notation: $\{subject: agent, action \}$ to describe a trust relationship between two entities, $subject$ and $agent$. Under this relationship, the trust is a function of uncertainty. The level of the trust is measured by a continuous real number. The trust value represents the uncertainty in a $subject$'s belief in an $agent$ for performing an $action$. Let $p\{subject: agent, action\}$ denote the probability that the $agent$ will perform the $action$ estimated by the $subject$. \cite{sun2006information} defines the trust-value $T\{subject: agent, action\}$ as:
	\begin{equation*}
	\scriptstyle
	T\{subject: agent, action\}=
	\begin{cases}
	1-H(p),& \quad for \mbox{ }0.5 \leq p \leq 1 \\
	H(p)-1,& \quad for \mbox{ }0 \leq p < 0.5 \mbox{ ,}
	\end{cases}
	\end{equation*}
	where $H(p)$ is the Shannon entropy. It should be noted that the trust value is calculated between two entities only and the trust value is an increasing function with $p$. The paper proposed a few methods in estimating the $p$ value, however, it is up to the application and circumstances in choosing the implementation strategies. 
	
	Saied et al.~\cite{saied2013trust} proposed to evaluate the trust value of node by taking into account parameters concerning its current context (service) and resource capabilities. The method considers all received $reports$ and past interactions. The computation of the trust value in the current context takes the centralized approach and consists four steps: 
	\begin{packed_enumc}
		\item selecting the set of nodes involving in trust value computations. 
		\item selecting the set $reports$ for each node in order to rate the trust level for each of the candidates. Since $reports$ come from different contexts with different requirements for the node capabilities other than the current one, a similarity metric is computed by defining a global contextual distance, which is related to both service difference and capacity difference. 
		\item each of the report is weighted according to its global contextual distance to the current context.
		\item the trust value $T_i$ for a node $i$ is then obtained as:
		\begin{equation*}
		T_i=\frac{1}{\sum_{j=1}^{n}w_{R_{ij}}} \sum_{j=1}^{n}(w_{R_{ij}} \cdot QR_j \cdot N_j ) \mbox{ ,}
		\end{equation*}
		where $w_{R_{ij}}$ is the weighting factor calculated in step 3, $QR_j$ is the quality recommendation for node $i$ issued by the $report$ $R_{ij}$, and $N_j$ is the score given by the requester node to the node for evaluating the offered service.
	\end{packed_enumc}
	
	Nitti et al.~\cite{nitti2014trustworthiness} integrated the social networking concepts into the IoT and estimated the trustworthiness of each node in both social and P2P scenarios. This trustworthiness is derived from P2P relationship between each node and its direct friends, which is referred as $T_{ij}$, the trust value of node $j$ estimated by node $i$. The trustworthiness is built on the basis of a node's own experience and on the basis of its friends. If two nodes are not friends, then the trust value is calculated by the trust value propagation through a chain of friendships. According to this new model, each node needs store and manage the feedback for calculating the trustworthiness level locally. Since the model is highly distributed, it is less sensitive to a single point of failure or infringement of the trust value. If two nodes are adjacent friends, the trust value is computed as follows:
	\begin{equation*}
	T_{ij}=(1-\alpha-\beta)R_{ij}+\alpha O_{ij}^{dir}+\beta O_{ij}^{ind} \mbox{ ,}
	\end{equation*}   
	where $R_{ij}$ is called the centrality of node $j$ with regard to node $i$. It represents the importance of the node $j$ from node $i$'s point of view. One way to quantify the importance is given by: $R_{ij}=\frac{|N_{ij}|}{|N_i|-1}$, where $|N_i|$ is the total number of transactions performed by node $i$, and $|N_{ij}|$ is the total number of transactions between node $i$ and node $j$. $O_{ij}^{dir}$ is estimated through node $i$'s own experience when interacting with node $j$. $O_{ij}^{ind}$ indicates evaluation coming from friends of the node $i$. For estimating trustworthiness between two nodes $i$ and $j$ that are not adjacent to each other, the trust value is calculated as the product of $T_{mn}$, where $m \sim n$ defines a friendship along the chain that connects node $i$ and node $j$.

	A trust-aware access control system for IoT (TACIoT) was proposed in \cite{bernabe2015taciot}. TACIoT extends the access control by taking trust values into account. TACIoT promotes a multidimensional trust model, which is a hierarchical and extensible approach to categorize multiple measurable properties and add new properties into the model. Four main dimensions included in the trust properties are: quality of service, security, reputation, and social relationship. For each trust property $p$ of a device $j$, its trust value $x_j^p$ is calculated as the product of observed property value and a \textit{Service-Important-Factor}, whose value depends on the relevance and importance of the service. The average rating of a device $\bar{a}$ is calculated as a weighted arithmetic mean based on paster interactions $n \in [1,...,N]$:
	\begin{equation*}
	\bar{a}^p_j=\frac{\sum_{n=1}^{N}x_{jn}^p * \exp^{-\frac{n}{S}}}{\sum_{n=1}^{N} \exp^{-\frac{n}{S}}} \mbox{ ,}
	\end{equation*}
	where $\exp^{-\frac{n}{S}}$ ensures that more weights are counted towards newest interactions. The expected value of a trust property $p$ for a peer IoT device $j$ depends on the average rating and the certainty value that indicates the reliability of the collected historical evidence. The overall trust value is recursively aggregated for each dimension with tunable weights for each branch and levels.

	\section{Conclusion}
IoT holds tremendous potential to improve our health, make our environment safer, boost productivity and efficiency, and conserve both water and energy. Cheating, however, is a real problem in the Internet of Things. The fundamental question that needs to be answered is how we can trust the validity of the data being generated in the first place. IoT therefore needs to improve its trustworthiness before it can be used to solve challenging economic and environmental problems tied to our social lives.
 
This paper presents an overview of the existing work on trust computing, access control models and systems in IoT. Access Control for IoT comprises the following three main types which include several sub-types. The first kind of access control model is role-based access control (RBAC), which is widely used in traditional networks. The second is credential-based access control (CBAC). In order to access various resources and data in this type of model, users require certain certificate information that falls into one of the following two types: Attribute-Based access control (ABAC) and Capability-Based access control (Cap-BAC). If a user has some special attributes in ABAC, it is possible to access a particular resource or piece of data. In Cap-BAC, a capability is a communicable, unforgeable rights markup, which corresponds to a value that uniquely specifies certain access rights to objects owned by subjects. The third kind of model is trust-based access control (TBAC) in which the trust computation models and trust management systems have been implemented successfully in commercial applications. There is also a rapidly growing literature dealing with topics of trust and reputation management for IoT. In addition, there are combinations of the aforementioned three methods. Some of the access control methods utilized to improve the security of the system include encryption and key management mechanisms.

The paper focuses on building an access control model and system based on trust computing, which is a new field of access control techniques that includes Access Control, Trust Computing, Internet of Things, network attacks, and cheating detection technologies. Because target access control systems can be very complex to manage, there has been substantial research in this domain, most of which has been related to attacks like self-promotion and ballot stuffing where a node falsely promotes its importance and boosts the reputation of a malicious node (by providing good recommendations) to engage in a collusion-style attack. The traditional trust computation model is inefficient in differentiating a participant object in IoT, which is designed to win trust by cheating. There is an urgent need to put forward more suitable and effective methods to ensure the security of IoT.

All in all, the paper presents an overview of the existing work on trust computing, access control models and systems in IoT. It not only summarizes the latest research progress, but also provides an understanding of current limitations and open issues. The paper is expected to provide useful guidelines for future research.

	

\Urlmuskip=0mu plus 1mu\relax
\bibliographystyle{splncs03}
\bibliography{myref1.bib}

\end{document}